\documentclass[%
 reprint, letterpaper,
showpacs,
preprintnumbers,
showkeys,
nofootinbib,
 amsmath,amssymb,
 aps, prd,
floatfix,
]{revtex4-1}
\usepackage{hyperref}
\usepackage[cp1250]{inputenc}
\newcommand{\ee}{\end{equation}}
\newcommand{\be}{\begin{equation}}
\newcommand{\ba}{\begin{array}}
\newcommand{\ea}{\end{array}}
\newcommand{\m}{M_{H^{\pm}}}
\newcommand{\g}{\,\mbox{GeV}}
\newcommand{\la}{\lambda_1}
\newcommand{\lb}{\lambda_2}
\newcommand{\lc}{\lambda_3}
\newcommand{\ld}{\lambda_4}
\newcommand{\lp}{\lambda_5}

\newcommand{\lczp}{\lambda_{345}}
\newcommand{\rg}{R_{\gamma\gamma}}
\newcommand{\rzg}{R_{Z\gamma}}
\newcommand{\rgt}{\widetilde{R}_{\gamma\gamma}}
\newcommand{\rzgt}{\widetilde{R}_{Z\gamma}}
\usepackage[pdftex]{graphicx}
\usepackage{amsmath}
\usepackage{amssymb}
\usepackage{url}
\usepackage{verbatim}
\begin{document}
\title{Diphoton rate in the Inert Doublet Model\\ with a 125~GeV Higgs boson}

\author{Bogumi\l a \'Swie\.zewska}
\email{Bogumila.Swiezewska@fuw.edu.pl}
\author{Maria Krawczyk}
\email{Maria.Krawczyk@fuw.edu.pl}

\affiliation{\textit{Faculty of Physics, University of Warsaw}\\ \textit{Ho\.za 69, 00-681 Warsaw, Poland}}
\preprint{IFT-9/2012}
\preprint{arXiv:1212.4100 [hep-ph]}
\date{15.01.2013}

\setlength\arraycolsep{2pt}

\begin{abstract}
An improved analysis of the diphoton decay rate of the Higgs boson in the Inert Doublet Model is presented together with a critical discussion of the  results existing in the literature. For a Higgs boson mass $M_h$ of $125\g$ and taking into account various constraints -- vacuum stability, existence of the Inert vacuum, perturbative unitarity, electroweak precision tests and the LEP bounds -- we find regions in the parameter space where the diphoton rate is enhanced. The resulting regions are confronted with the allowed values of the Dark Matter mass. We find that a significant enhancement in the two-photon decay of the Higgs boson is only possible for constrained values of the scalar couplings $\lambda_3 \sim hH^+H^-$, $\lczp \sim h HH$ and the masses of the charged scalar and the Dark Matter particle. The enhancement above 1.3 demands that the masses of $H^{\pm}$ and $H$  be less than $135\g$  (and above $62.5\g$) and $-1.46<\lambda_3,\lambda_{345} <-0.24$. In addition, we analyze the correlation of the diphoton and $Z\gamma$ rates.
\end{abstract}

\maketitle
\section{Introduction}
Recently a Higgs-like boson was discovered at the LHC~\cite{atlas:2012,*cms:2012}. Although most of the  measurements of its properties are in agreement with the hypothesis of the Standard Model (SM) Higgs particle, there may be some indications that the discovered boson is not a SM Higgs boson. For example, the signal strength in the decay channel $h\to\gamma\gamma$, $\rg$, which is proportional to $\textrm{Br}(h\to \gamma \gamma)$, is equal to  $1.8\pm 0.3$~\cite{Atlas:12-2012}. This can be accounted for in the framework of Two Higgs Doublet Models (2HDM), in particular in the Inert Doublet Model (IDM).

The diphoton decay rate in the IDM was considered in Refs.~\cite{Ma:2007, Posch:2010, Arhrib:2012, Chang:2012}. In the parameter region studied in Ref.~\cite{Ma:2007} no enhancement was found, while in Ref.~\cite{Posch:2010} the possibility of modifying the total decay width of the Higgs boson due to the invisible decays into Dark Matter (DM) was not taken into account. In Refs.~\cite{Arhrib:2012, Chang:2012}  the entire parameter space was not investigated; as  the mass parameter of the potential was taken with only one sign, the DM particle was assumed to be lighter than the Higgs boson and the mass of DM was constrained ($M_H<150\g$). 
The diphoton decay rate was also considered in the context of the electroweak phase transition in Ref.~\cite{Cline:2012}.

We present an independent analysis of the diphoton Higgs decay mode in the IDM which improves the   points mentioned above and makes use of the recent experimental data~\cite{atlas:2012,cms:2012}.
 We also study the correlation between the $\gamma\gamma$ and $Z\gamma$ Higgs boson decay rates.
The paper is organized as follows. In Sec.~\ref{model} we briefly review the model, describe the constraints taken into account and present the method of the analysis. In Sec.~\ref{modyfikacje} possible sources of modifications of the diphoton Higgs decay rate in the IDM with respect to the SM are discussed. The resulting constraints for masses and  self-couplings as well as the results for the $Z\gamma$ rate are presented in Sec.~\ref{wyniki}, and a short summary can be found in Sec.~\ref{sum}. Appendix A contains a derivation of the conditions for $\rg>1$, while in Appendix B formulas for all Higgs boson decay widths are given.


\section{Setup of the analysis}\label{model}

\subsection{Model}

We consider the IDM~\cite{Ma:1978,Barbieri:2006, Ma:2007}, which is a 2HDM with two $\textrm{SU}(2)$ doublets $\phi_S$, $\phi_D$ with hypercharge $Y=1$ and the following potential:
\be\label{pot}\renewcommand{\arraystretch}{1.5}
\begin{array}{rcl}
V&=&-\frac{1}{2}\left[m_{11}^{2}(\phi_{S}^{\dagger}\phi_{S})+m_{22}^{2}(\phi_{D}^{\dagger}\phi_{D})\right]\\*
&&+\frac{1}{2}\left[\lambda_{1}(\phi_{S}^{\dagger}\phi_{S})^{2}+\lambda_{2}(\phi_{D}^{\dagger}\phi_{D})^{2}\right]\\*
&&+\lambda_{3}(\phi_{S}^{\dagger}\phi_{S})(\phi_{D}^{\dagger}\phi_{D})+\lambda_{4}(\phi_{S}^{\dagger}\phi_{D})(\phi_{D}^{\dagger}\phi_{S})\\*
&&+\frac{1}{2}\lambda_{5}\left[(\phi_{S}^{\dagger}\phi_{D})^{2}+(\phi_{D}^{\dagger}\phi_{S})^{2}\right].\\*
\end{array}
\ee
The parameters $m_{11}^{2}$, $m_{22}^{2}$, and $\la\ldots\ld$ are real numbers and without loss of generality we take $\lp<0$~\cite{Krawczyk:2010,  Krawczyk:2004sym, Branco:1999}. The~potential $V$ is invariant under a $\mathbb{Z}_2$-type symmetry transformation, called $D$, which changes the sign of the $\phi_D$ doublet and leaves all other fields unchanged.

We consider a $D$-symmetric vacuum state (called the Inert vacuum), which corresponds to the following vacuum expectation values  \renewcommand{\arraystretch}{.7}$\langle\phi_{S}\rangle=\left(\begin{array}{c}0\\v/\sqrt{2}\\\end{array}
\right)$, $\langle\phi_{D}\rangle=0$. The masses of the physical scalars read:\renewcommand{\arraystretch}{1.5}
$$
\begin{array}{rcl}
M_{h}^{2}&=&m_{11}^{2}=\la v^{2},\\
M_{H^{\pm}}^{2}&=&\frac{1}{2}(\lc v^{2}-m_{22}^{2}),\\
M_{A}^{2}&=&\frac{1}{2}(\lczp^{-}v^{2}-m_{22}^{2})=\m^2+\frac{1}{2}\lambda_{45}^- v^2,\\
M_{H}^{2}&=&\frac{1}{2}(\lczp v^{2}-m_{22}^{2})=\m^2+\frac{1}{2}\lambda_{45} v^2,\\
\end{array}
$$
where $\lczp=\lc+\ld+\lp,\;\lczp^-=\lc+\ld-\lp,\; \lambda_{45}^-=\ld-\lp,\;\lambda_{45}=\ld+\lp $. The parameters $\lc$ and $\lczp$ are proportional to the $hH^+H^-$ and $hHH$ couplings, respectively. The only parameter of the potential, that is absent in these formulas is the parameter $\lambda_2$, related to the quartic self-couplings among the scalars, e.g. $HHHH$ or $H^+H^-H H$.

Since the $\phi_S$ doublet solely takes part in the Spontaneous Symmetry Breaking (SSB), only $h$ is a Higgs boson. The remaining scalars are often called inert scalars\footnote{They are also called dark scalars ($D$-scalars).} as they do not contribute to SSB.

In the IDM  the Yukawa interactions are set to  Type~I, with the $\phi_S$ doublet  coupled to fermions. Thus $h$ is SM-like, i.e., it couples to fermions (at the tree level)  just like the SM Higgs and  to the gauge bosons as well. We assume that it corresponds to the boson discovered at the LHC in 2012~\cite{atlas:2012, *cms:2012} and set $M_h=125\g$.

Since the $D$ symmetry  is exact in the IDM, it renders the lightest neutral $D$-odd particle stable providing a good DM candidate~\cite{Dolle:2009, *LopezHonorez:2006, *LopezHonorez:2007,  *Sokolowska:2011}. Without loss of generality we  assume that the DM particle is the $H$ scalar, so $\lambda_{45}<0$.\footnote{
It does not lead to the limitation of the mass of the Higgs boson, as the DM particle is
the lightest of the $D$-odd scalars, rather than of all scalars, as was assumed in Ref.~\cite{Arhrib:2012}.}

\subsection{Constraints}\label{warunki}

In our analysis we took into account the following constraints~\cite{Swiezewska:2012}
\begin{description}
\item[Vacuum stability] For a stable vacuum state to exist it is necessary that the potential $V$ is bounded from below, which leads to~\cite{Ma:1978}
\begin{displaymath}
\la>0,\quad\lb>0,\quad\lc+\sqrt{\la\lb}>0,\quad\lczp+\sqrt{\la\lb}>0.
\end{displaymath}
\item[Perturbative unitarity] For the theory to be perturbatively unitary it is required that the eigenvalues $\Lambda_i$ of the high-energy scattering matrix fulfill the condition $|\Lambda_i|<8\pi$,~\cite{Kanemura:1993,*Akeroyd:2000,Swiezewska:2012}. In particular, we got $\la,\lb<8.38$ and $\lc<16.2$.
\item[Existence of the Inert vacuum] The Inert vacuum can be realized  only if the following conditions are fulfilled~\cite{Krawczyk:2010, praca-mag}:
\begin{displaymath}
M_h^2,\, M_H^2,\,M_A^2,\,\m^2\geqslant0,\quad\frac{m_{11}^2}{\sqrt{\la}}>\frac{m_{22}^2}{\sqrt{\lb}}.
\end{displaymath}

From the existence of the Inert vacuum and the Higgs boson with mass $M_h=125\g$, and unitarity bounds on $\lb$,  follows a bound on $m_{22}^2$~\cite{Swiezewska:2012}:
\begin{equation}\label{m22bound}
m_{22}^2\lesssim 9\cdot10^4\g^2.
\end{equation}
\item[$H$ as DM candidate] We assume that $H$ is the DM candidate, so $M_H<M_A,\m$. Studies of the DM in the IDM~\cite{Dolle:2009, *Sokolowska:2011, *LopezHonorez:2006, *LopezHonorez:2007} show that if $H$ is to account for the observed  relic density of DM, it should have a mass in one of three regions:  $M_H<10\g$,  $40\g<M_H<80\g$, or  $M_H>500\g$. We will not impose these bounds from the very beginning, but we will discuss the consistency of our results with these constraints.
\item[Electroweak Precision Tests (EWPT)] We demand that the values of the $S$ and $T$ parameters calculated in the IDM (using formulas from Ref.~\cite{Barbieri:2006}) lie within $2\sigma$ ellipses in the $S,T$ plane, with the following central values~\cite{Nakamura:2010}: $S=0.03\pm0.09$, $T=0.07\pm0.08$, with correlation equal to 87\%.
\item[LEP] We use the LEPI and LEPII bounds on the scalar masses~\cite{Gustafsson:2009,Gustafsson:2010}
\begin{eqnarray}
&\m +M_H>M_{W},\quad \m+ M_A>M_W,\nonumber\\
&M_H+M_A >M_Z,\quad  \m>70\g \nonumber
\end{eqnarray}
and exclude the region where simultaneously $M_H< 80\g$, $M_A< 100\g$ and $M_A - M_H> 8\g$.
\end{description}
We will refer to the set of the conditions described above as ``the constraints'' for simplicity.
\subsection{Method of the analysis}
We randomly scan the parameter space of the IDM, taking into account the constraints and letting the parameters vary in the following regimes:
\begin{equation}
\begin{array}{rcccl}
&&M_h&=&125\g,\nonumber\\[-4pt]
70\g&\leqslant&\m&\leqslant&800\;(1400)\g,\nonumber\\[-4pt]
0&<&M_A&\leqslant&800\;(1400)\g,\nonumber\\[-4pt]
0&<&M_H&<&M_A,\m,\nonumber\\[-4pt]
-25\cdot10^4\;(-2\cdot10^6)\g^2&\leqslant&m_{22}^2&\leqslant&9\cdot10^4\g^2,\nonumber\\[-4pt]
0&<&\lb&\leqslant &10.\nonumber\\
\end{array}
\end{equation}
 The allowed region in the parameter space depends on the choice of the minimal value of $m_{22}^2$, which is not constrained. We consider two regimes for $m_{22}^2$. For the wider of the two, larger masses of dark scalars are allowed, up to $1400\g$  (values in brackets).

In the parameter space fulfilling the constraints we analyze the possible values of $\rg$ and $\rzg$.
\section{$\rg$}\label{modyfikacje}

To study the diphoton rate observed at the LHC we define the quantity $\rg$  as follows~\cite{Arhrib:2012}:
\begin{align}\label{rgg}
R_{\gamma \gamma}&:=\frac{\sigma(pp\to h\to \gamma\gamma)^{\textrm{IDM}}}{\sigma(pp\to h\to \gamma\gamma)^{\textrm  {SM}}}\nonumber\\*
&\approx\frac{\left[\sigma(gg\to h) \textrm{Br}(h\to\gamma\gamma)\right]^{\textrm {IDM}}}{\left[\sigma(gg\to h) \textrm{Br}(h\to\gamma\gamma)\right]^{\textrm {SM}}}
=\frac{\textrm{Br}(h\to\gamma\gamma)^{\textrm {IDM}}}{\textrm{Br}(h\to\gamma\gamma)^{\textrm {SM}}}.
\end{align}
Above we used the fact that the gluon fusion is the dominant channel of Higgs production. Moreover, in the IDM $\sigma(gg\to h)^{\textrm {IDM}}=\sigma(gg\to h)^{\textrm {SM}}$, so $\rg$ reduces to the ratio of branching ratios.

In the IDM this ratio can be modified with respect to the SM, since the charged scalar exchanged in loops gives an extra contribution to the $h\to\gamma\gamma$ amplitude~\cite{Ma:2007, Posch:2010, Arhrib:2012}. In addition, the total decay width of the Higgs boson can be modified due to the existence of invisible decay channels $h\to HH$ and $h\to AA$~\cite{Ma:2007, Arhrib:2012}. In different regions of parameters, different  effects dominate.

Many channels contribute to the total decay width of the Higgs boson $h$. The most important ones for a mass of $M_h=125\g$ are $b\overline{b}$, $c\overline{c}$, $\tau^+\tau^-$, $ZZ^*$, $WW^*$, $\gamma\gamma$, $Z\gamma$, $gg$, $HH$, and $AA$.  To compute the decay widths we used the formulas from Refs.~\cite{Djouadi:2005, *Djouadi:2005sm, Djouadi:1995,Beringer:2012, Chen:2013}. For completeness they are summarized in Appendix~\ref{dec}. The partial widths of the tree-level $h$ decays into SM particles, and  the loop-mediated decay into $gg$ in the IDM are equal to the corresponding ones in the SM.

The only decay rate of $h$ -- apart from $h\to\gamma\gamma$ -- that is modified in the IDM with respect to the SM is that for the $h\to Z\gamma$ process, $\rzg$. It is defined in the same way as $\rg$
$$
R_{Z \gamma}=\frac{\textrm{Br}(h\to Z\gamma)^{\textrm {IDM}}}{\textrm{Br}(h\to Z\gamma)^{\textrm {SM}}}
$$
and $\Gamma(h\to Z\gamma)^{\textrm {IDM}}$ is given by the formula~(\ref{rzg-wzor}). We discuss some results for $\rzg$ in Sec.~\ref{rzg}.

\begin{widetext}

\begin{figure}[h]
\centering
\includegraphics[width=.33\textwidth]{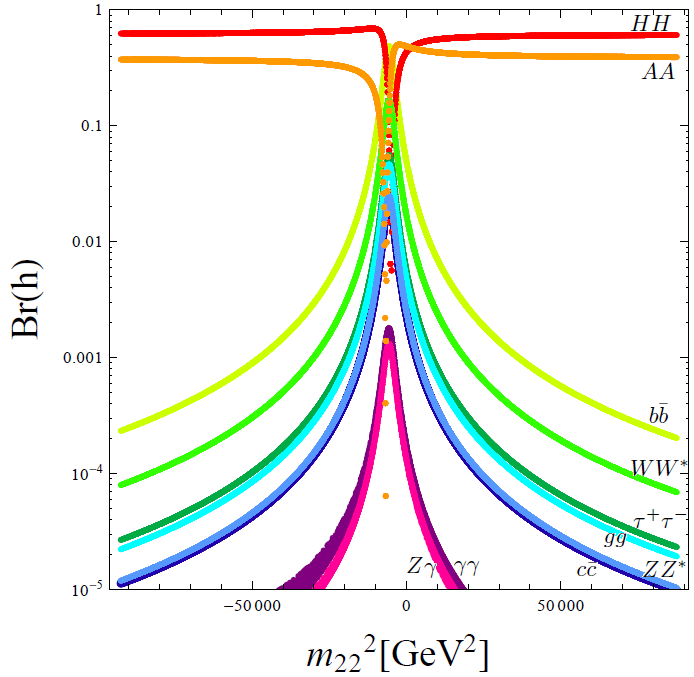}
\includegraphics[width=0.33\textwidth]{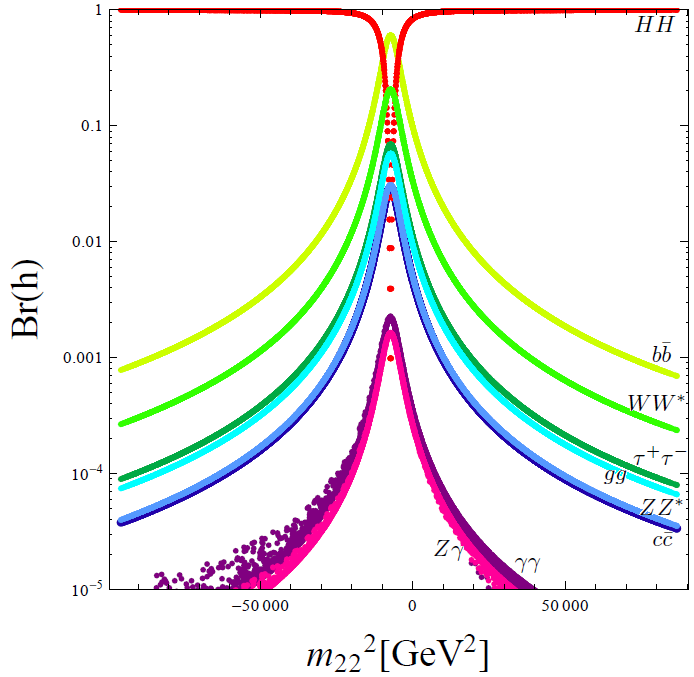}
\includegraphics[width=0.33\textwidth]{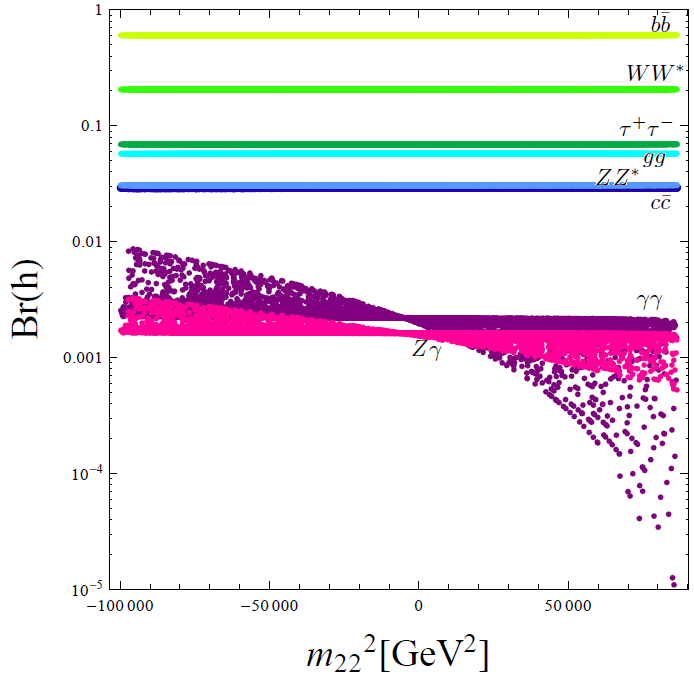}
\caption{Branching ratios for $h$ with mass 125 GeV. Left panel: decay channels $h\to HH$ and $h\to AA$ are open  ($M_H=50\g$, $M_A=58\g$), middle panel: $h\to HH$ open ($M_H=60\g, M_A>63\g$), right panel: no invisible $h$ decay channels allowed ($M_H=75\g, M_A>M_H$).\label{fig:br}}
\end{figure}

\end{widetext}

In Fig.~\ref{fig:br} the  branching ratios of $h$ are presented as functions of $m_{22}^2$. Three different cases are considered: decay channels $h\to AA$ and $h\to HH$ are open (with $M_H=50\g$, $M_A=58\g$: left panel), $h\to AA$ is closed and $h\to HH$ is open ($M_H=60\g$, $M_A>63\g$:  middle panel), and both $h\to AA$ and $h\to HH$ are closed ($M_H=75\g$, $M_A>M_H$: right panel).

It appears that when the invisible decay channels $h\to HH$ and $h\to AA$ are open  ($M_H<M_h/2$, $M_A<M_h/2$), their partial widths $\Gamma(h\to HH)$, $\Gamma(h\to AA)$ dominate over the  partial widths of decays into SM particles. It will be shown, that in these cases the total decay width of the Higgs boson is so big that  $\textrm{Br}(h\to\gamma\gamma)^{\textrm{IDM}}<\textrm{Br}(h\to\gamma\gamma)^{\textrm{SM}}$ always.

When $M_H>M_h/2$ (and therefore $M_A>M_h/2$ as well) the invisible decay channels are closed. Then all the branching ratios are constant (Fig.~\ref{fig:br}, right panel), with the exception of  $\textrm{Br}(h\to\gamma\gamma)$ and $\textrm{Br}(h\to Z\gamma)$, which vary significantly with $m_{22}^2$. We will analyze this case below.

If the decay channels $h\to HH$ and $h\to AA$ are kinematically closed, the total width of $h$ is barely modified with respect to the SM case, since the branching ratios of $h\to \gamma \gamma$ and $h\to Z \gamma$, which are the only processes that receive contributions from dark scalars, are of the order of $10^{-3}$. Thus $\rg$ and $\rzg$ [Eq.~(\ref{rgg})] reduce to the ratios of  the partial widths in the IDM to the ones in the SM, namely
\be\label{rggfalka}
\widetilde{R}_{\gamma \gamma}=\frac{\Gamma(h\to\gamma\gamma)^{\textrm {IDM}}}{\Gamma(h\to\gamma\gamma)^{\textrm {SM}}},\quad \widetilde{R}_{Z \gamma}=\frac{\Gamma(h\to Z\gamma)^{\textrm {IDM}}}{\Gamma(h\to Z\gamma)^{\textrm {SM}}}.
\ee
In the IDM, the partial decay width of the Higgs boson to $\gamma \gamma$  is (approximately) given by~\cite{Djouadi:2005, *Djouadi:2005sm, Ma:2007, Posch:2010, Arhrib:2012}
\begin{eqnarray}
\Gamma(h\to\gamma\gamma)^{\textrm{IDM}}&=&\frac{G_F\alpha^2M_h^3}{128\sqrt{2}\pi^3}\bigg | \underbrace{\frac{4}{3}  A_{1/2}\bigg(\frac{4M_t^2}{M_h^2}\bigg)+ A_1\bigg(\frac{4M_W^2}{M_h^2}\bigg)}_{\mathcal{M}^{\textrm{SM}}}\nonumber\\
&&+\underbrace{\frac{2\m^2+m_{22}^2}{2\m^2}A_0\bigg(\frac{4\m^2}{M_h^2}\bigg)}_{\delta\mathcal{M}^{\textrm{IDM}}}\bigg |^2,\nonumber
\end{eqnarray}
where $\mathcal{M}^{\textrm{SM}}$ denotes the contribution from the SM and $\delta\mathcal{M}^{\textrm{IDM}}$ is the extra contribution present in the IDM, $\mathcal{M}^{\textrm{IDM}}=\mathcal{M}^{\textrm{SM}}+\delta\mathcal{M}^{\textrm{IDM}}$.\footnote{Above we do not include the contributions from the bottom- and charm-quark loops as well as from the $\tau$ loop, as we have checked that they are negligible. We take $M_W=80.399\g$ and $M_t=173\g$ from the Particle Data Group analysis~\cite{Beringer:2012}.}

The form factors are defined as follows~\cite{Spira:1997}:
\begin{eqnarray}
A_0(\tau)&=&-\tau[1-\tau f(\tau)],\nonumber\\
A_{1/2}(\tau)&=&2\tau[1+(1-\tau)f(\tau)],\nonumber\\
A_1(\tau)&=&-[2+3\tau+3\tau(2-\tau)f(\tau)]\nonumber
\end{eqnarray}
and
\begin{displaymath}
f(\tau)=\begin{cases}
\arcsin^2\big(\frac{1}{ \sqrt{\tau} }\big) & \textrm{for } \tau\geqslant 1,\\[4pt]
-\frac{1}{4}\Big[\log\Big(\frac{1+\sqrt{1-\tau}}{1-\sqrt{1-\tau}}\Big)-i\pi\Big]^2  & \textrm{for } \tau<1.
\end{cases}
\end{displaymath}

The $h \to \gamma \gamma$ enhancement is of interest to us, so we consider the inequality $\rgt>1$, which corresponds to :
\be\label{rggt1}
|\mathcal{M}^{\textrm{SM}}+\delta\mathcal{M}^{\textrm{IDM}}|^2>|\mathcal{M}^{\textrm{SM}}|^2,
\ee
where $\mathcal{M}^{\textrm{SM}}$ is fixed for $M_h=125\g$.\footnote{If the contributions from light quarks are neglected, $\mathcal{M}^{\textrm{SM}}$ is real, but  we treat it as a complex number to keep the reasoning general.} 
  The inequality~(\ref{rggt1}) can be solved analytically (the~full derivation is given in Appendix~\ref{wypr}), giving the result that $\rg>1$ is possible only when
\begin{align}\label{war-m22}
&m_{22}^2<-2\m^2\quad
\textrm{or}\nonumber\\*[4pt]
&m_{22}^2>\frac{M_h^2\,\textrm{Re}\left(\mathcal{M}^{\textrm{SM}}\right) }{1-\left(\frac{2 \m}{M_h}\right)^2\arcsin^2\left(\frac{M_h}{2 \m}\right)}-2\m^2.
\end{align}
These two conditions correspond to two possible cases: when the contribution of the charged scalar loop interferes either constructively or destructively with the SM contribution. In the latter case the contribution from the charged scalar has to be at least twice as big as the SM term~\cite{Posch:2010}.
Since both of the functions in Eq.~(\ref{war-m22}) are monotonic (with respect to $\m$) we can get overall bounds on $m_{22}^2$ with the use of the LEPII  bound on the mass of the charged scalar. Substituting $\m=70\g$ and $M_h=125\g$ into these bounds yields $m_{22}^2<-9.8\cdot10^3\g^2$ or $m_{22}^2\gtrsim1.8\cdot 10^5\g^2$. Taking into account the bound~(\ref{m22bound}) we are left with the only  option (constructive interference):
\be\label{wyn}
m_{22}^2<-9.8\cdot10^3\g^2.
\ee

The conditions~(\ref{war-m22}) can be translated into conditions for the $hH^+H^-$ coupling ($~\lc$) with the use of the expression for the mass of the charged scalar, $M_{H^{\pm}}^{2}=\frac{1}{2}(\lc v^{2}-m_{22}^{2})$, giving the condition $\lc<0$.


\section{Results}\label{wyniki}
\subsection{$\rg$}\label{rg}

In this section we present the regions in the parameter space allowed by the constraints (Sec.~\ref{warunki}) and the condition $\rg>1$. Points with $\rg<1$ are displayed in Figs. 2-5 in dark green/gray and with $\rg>1$ in light green/gray.

\begin{figure}[h]
\centering
\includegraphics[width=.4\textwidth]{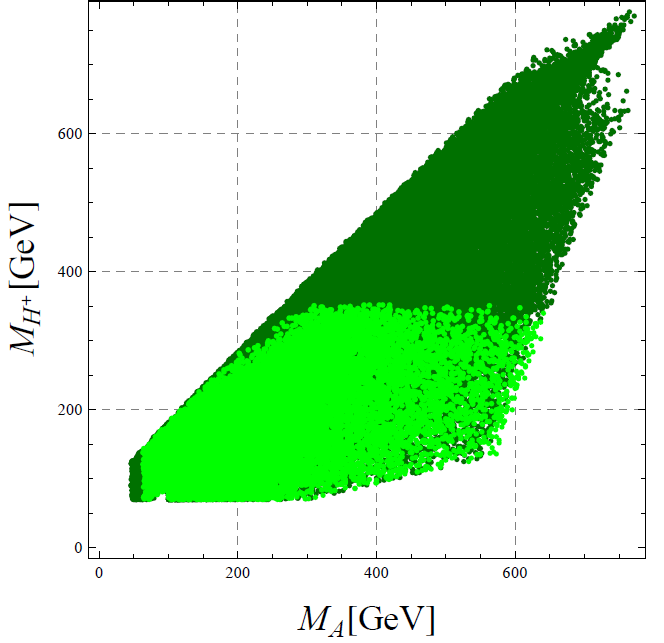}\\
\includegraphics[width=0.4\textwidth]{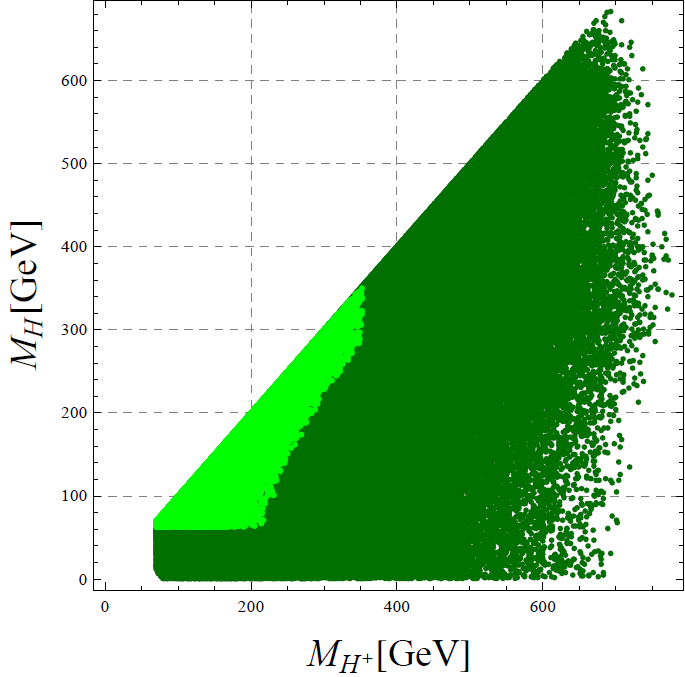}
\caption{Regions in the $(M_A,\m)$ (upper panel) and $(\m,M_H)$ (lower panel) planes allowed by the constraints for $-25\cdot10^4\g^2 \leqslant m_{22}^2 \leqslant 9\cdot10^4\g^2$. Points with $\rg<1$ ($\rg>1$) are displayed in dark green/gray (light green/gray). Note that in the upper panel the light green region overlaps the dark green one, since for given values of $\m$ and $M_A$ the mass of $H$ can vary, rendering the $h\to HH$ channel open or closed.
\label{fig:masy}}
\end{figure}

In Fig.~\ref{fig:masy}
 the regions of masses allowed in the IDM by the constraints for the narrow $m_{22}^2$ range (Sec. II.C) are presented with the regions where the enhancement in the $h\to \gamma\gamma$ channel is singled out. We have found that the $\rg$ enhancement is only possible when $M_H>M_h/2$ and $M_A>M_h/2$. It means that the partial widths of invisible decays increase the total width of the Higgs boson so much that the enhancement with respect to the SM case is impossible (this is in agreement with the results of Ref.~\cite{Arhrib:2012}).\footnote{Figure~\ref{fig:masy} (upper panel) differs from Fig.~1 (left panel) of Ref.~\cite{Arhrib:2012} as in Ref.~\cite{Arhrib:2012}  the DM particle ($H$) was assumed to be lighter than the Higgs boson ($M_H<150\g$), which decreased the size of the allowed region and also constrained the values of $m_{22}^2$, which resulted in tighter upper bounds on the masses of $H^{\pm}$ and $A$.}

\begin{figure}[h]
\centering
\includegraphics[width=0.4\textwidth]{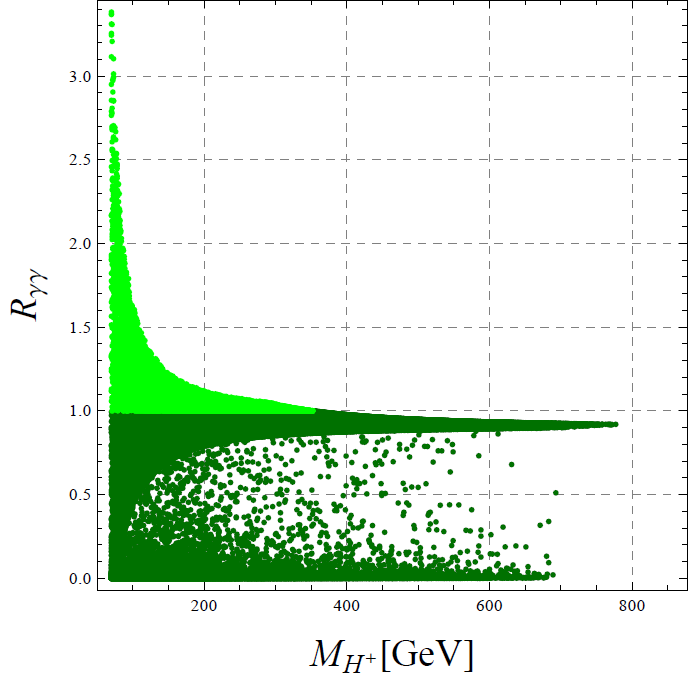}\\
\includegraphics[width=.4\textwidth]{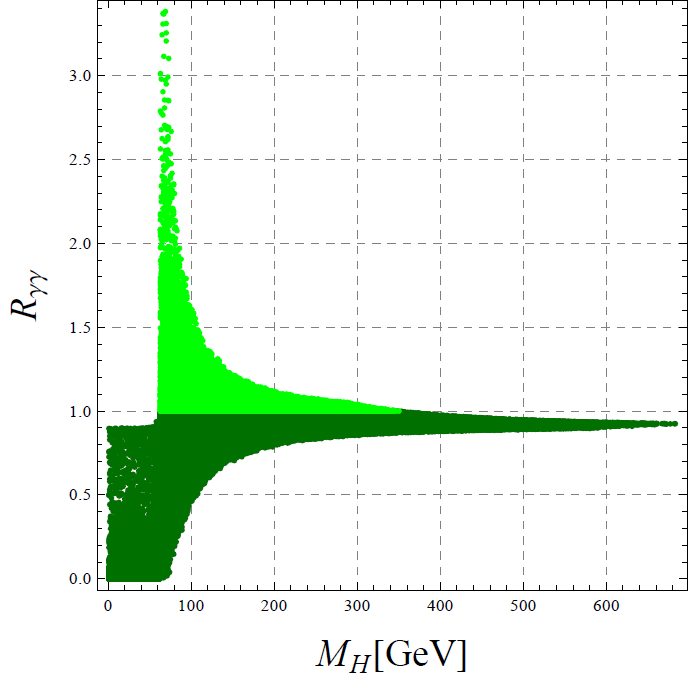}
\caption{Values of $\rg$ allowed by the constraints for $-25\cdot10^4\g^2 \leqslant m_{22}^2 \leqslant 9\cdot10^4\g^2$ as a function of masses: $\m$ (upper panel) and $M_H$ (lower panel). Points with $\rg<1$ ($\rg>1$) are displayed in dark green/gray (light green/gray).\label{fig:mr}}
\end{figure}

The signal in the $h\to\gamma\gamma$ channel can be enhanced with respect to the SM up to around 3.4 times, which can be inferred from Fig.~\ref{fig:mr}, where the dependence of $\rg$ on   $M_H$ and $\m$ is presented.

\begin{figure}[h]
\centering
\includegraphics[width=0.4\textwidth]{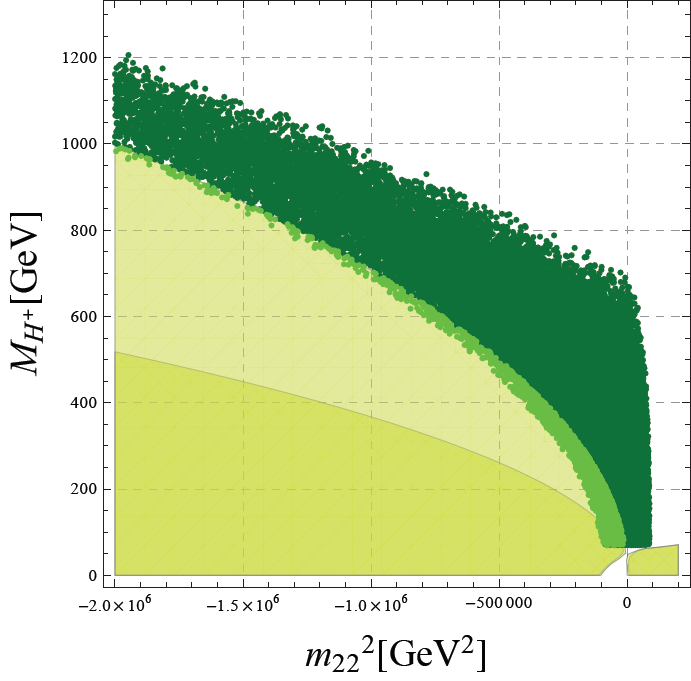}\\
\includegraphics[width=0.4\textwidth]{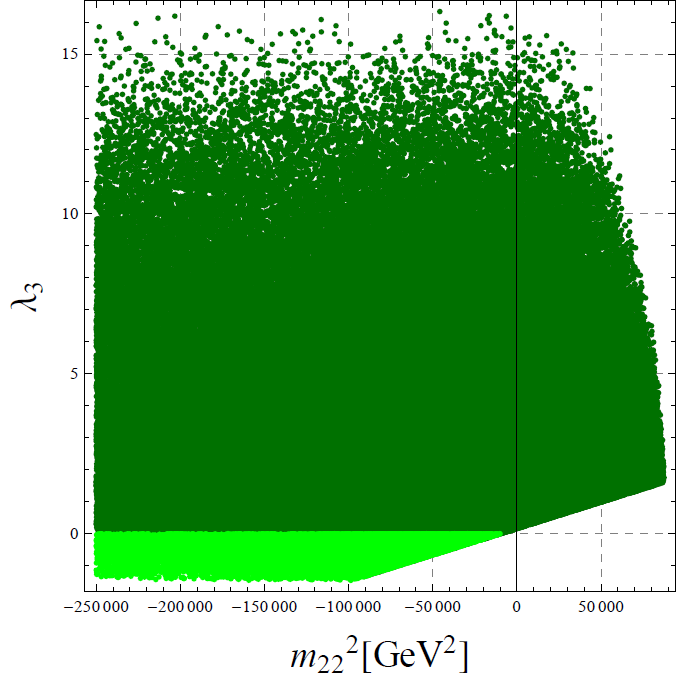}
\caption{Upper panel: Regions where $\rgt>1$ ($\rgt>1.3$) [see Eq.~(\ref{rggfalka})] are displayed as light (dark) shaded. Results of the scan for the region $-2\cdot10^6\g^2 \leqslant m_{22}^2 \leqslant 9\cdot10^4\g^2$ with $\rg<1$ ($\rg>1$) are displayed in dark (light) green/gray. There is an overlap of the light green and  shaded regions and also of the light and dark green regions. Lower panel: Region allowed by the constraints for $-25\cdot10^4\g^2 \leqslant m_{22}^2 \leqslant 9\cdot10^4\g^2$ in the $(m_{22}^2,\lc)$ plane. Points with $\rg<1$ ($\rg>1$) are displayed in dark green/gray (light green/gray).
\label{fig:m22}}
\end{figure}

Figures~\ref{fig:masy} and~\ref{fig:mr} seem to suggest that $\rg>1$ is only possible for $\m\lesssim 350\g$ (compare with the bound $\m\lesssim200\g$ from Ref.~\cite{Arhrib:2012}). However, this is not the case. If we allow for a wider $m_{22}^2$ range, then we get a larger $\m$ for which $\rg>1$.\footnote{In Ref.~\cite{Arhrib:2012} $m_{22}^2$ (or $\mu_2^2=- 1/2 m_{22}^2$) was limited by setting $M_H<150\g$.} This fact is illustrated in the upper panel of Fig.~\ref{fig:m22},
 where the results of the scan with a wider range of $m_{22}^2$, are presented. It can be seen that $\rg>1$ for $\m$ up to $1\,\textrm{TeV}$.  Regions fulfilling $\rgt>1$ (light shaded region) and $\rgt>1.3$ (dark shaded region) are also shown. 

It can be seen from Fig.~\ref{fig:mr} (upper panel) that substantial enhancement  ($\rg> 1.3$) appears for a relatively light charged scalar, $\m\lesssim135\g$. Moreover, $\rg>1.3$
is only possible in a region $m_{22}^2\gtrsim-1.3\cdot10^5\g^2$ (see Fig.~\ref{fig:m22}, upper panel), and hence the upper bound on $\m$ does not change if we allow for a very big negative $m_{22}^2$.

Figure~\ref{fig:mr} shows that if $\rg< 1.3$  the DM particle also has to be light: $M_H\lesssim 135\g$.  So we conclude that a substantial enhancement is only possible for
$$
\ba{rcccl}
62.5\g&<&M_H&<&135 \g,\\[-4pt]
70\g&<&\m&<&135 \g.\\
\ea
$$

This reasoning is general and will give upper bounds on $\m$ and $M_H$ if the  enhancement of $h\to \gamma \gamma$ decay with respect to the SM,  $\rg>1$, is definitely confirmed by data.

Stronger constraints could be obtained for the DM particle if additional data were included in the analysis, e.g., the WMAP data on the relic abundance of the DM. For the sake of clarity, we leave the detailed analysis of the Inert DM properties in light of the $\rg$, WMAP, and XENON100 data for a separate study~\cite{Krawczyk:2013}.

From the upper panel of Fig.~\ref{fig:m22} it is also visible that an additional region fulfilling $\rg>1$ would be allowed if  $m_{22}^2$ could be greater than $9\cdot10^4\g^2$.\footnote{In Ref.~\cite{Arhrib:2012} the conditions determining the existence of the Inert vacuum were not taken into account, 
so a region with $\rg>1$ and $m_{22}^2>0$ would appear in that analysis: only by choosing the maximal value of DM particle's mass, $M_H$, was this unphysical 
region fortunately avoided.}
However this  is not allowed for the Inert vacuum (\ref{war-m22}).
 This shows the important  role of the conditions determining the existence of the Inert vacuum in constraining the parameter space and the need for taking them into account.

In the lower panel of Fig.~\ref{fig:m22} we present the allowed region in the $(m_{22}^2,\ \lc)$ plane ($\lambda_3 \sim h H^+H^-$ coupling) 
and confirm the conclusion of Ref.~\cite{Arhrib:2012} that the enhanced diphoton production rate is possible only for $\lc<0$.

\begin{figure}[h]
\centering
\includegraphics[width=.4\textwidth]{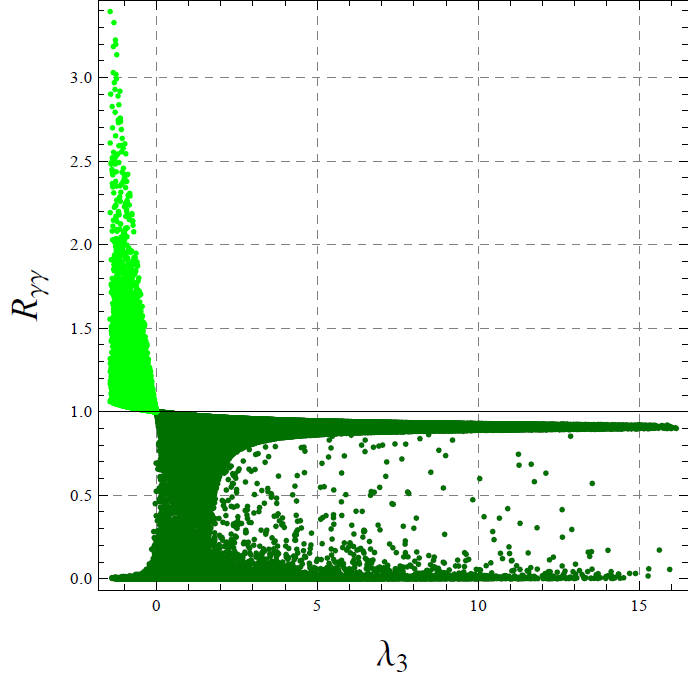}\\
\includegraphics[width=0.4\textwidth]{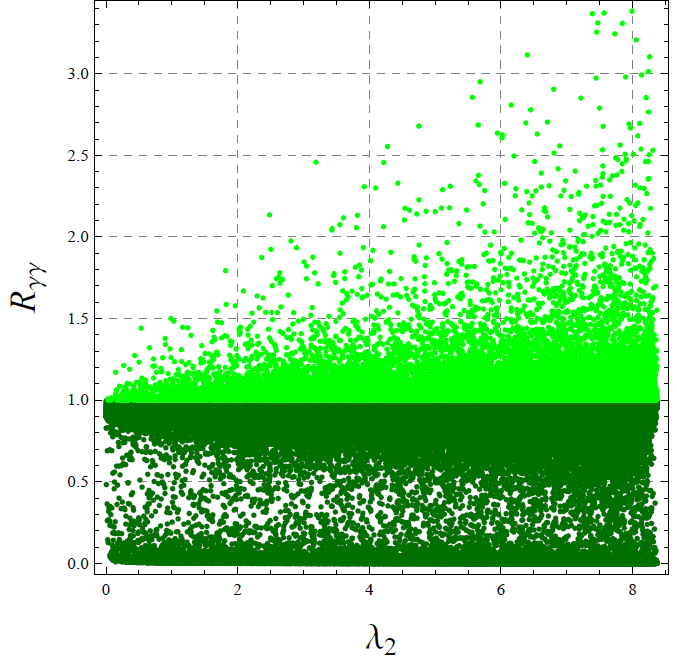}
\caption{Values of $\rg$ allowed by the constraints for $-25\cdot10^4\g^2 \leqslant m_{22}^2 \leqslant 9\cdot10^4\g^2$ as a function of the couplings $\lc$ (upper panel) and $\lb$ (lower panel). Points with $\rg<1$ ($\rg>1$) are displayed in dark green/gray (light green/gray).
\label{fig:lr}}
\end{figure}

$\rg$ as a function of the couplings $\lc$ and $\lb$ is shown in the upper and lower panels of Fig.~\ref{fig:lr}, respectively. It can be seen once more that $\rg>1$ for $\lc<0$ and as a consequence, since $M_H<\m$,  $\lczp<0$ with quite stringent lower bounds on both of the couplings,
$$
\lc,\lczp>-1.47.
$$
If in addition $\rg>1.3$, then
$$
 -1.46<\lc,\lczp< -0.24.
$$
On the contrary, $\rg>1$ is possible for all values of $\lambda_2$ (lower panel of Fig.~\ref{fig:lr}).\footnote{It does not agree with the observation of Ref.~\cite{Posch:2010} which stated that $\rg>1$ for $-\lc\gg 1$ and only for big values of $\lb$.}

\subsection{$\rzg$}\label{rzg}

\begin{figure}[h]
\centering
\includegraphics[width=.4\textwidth]{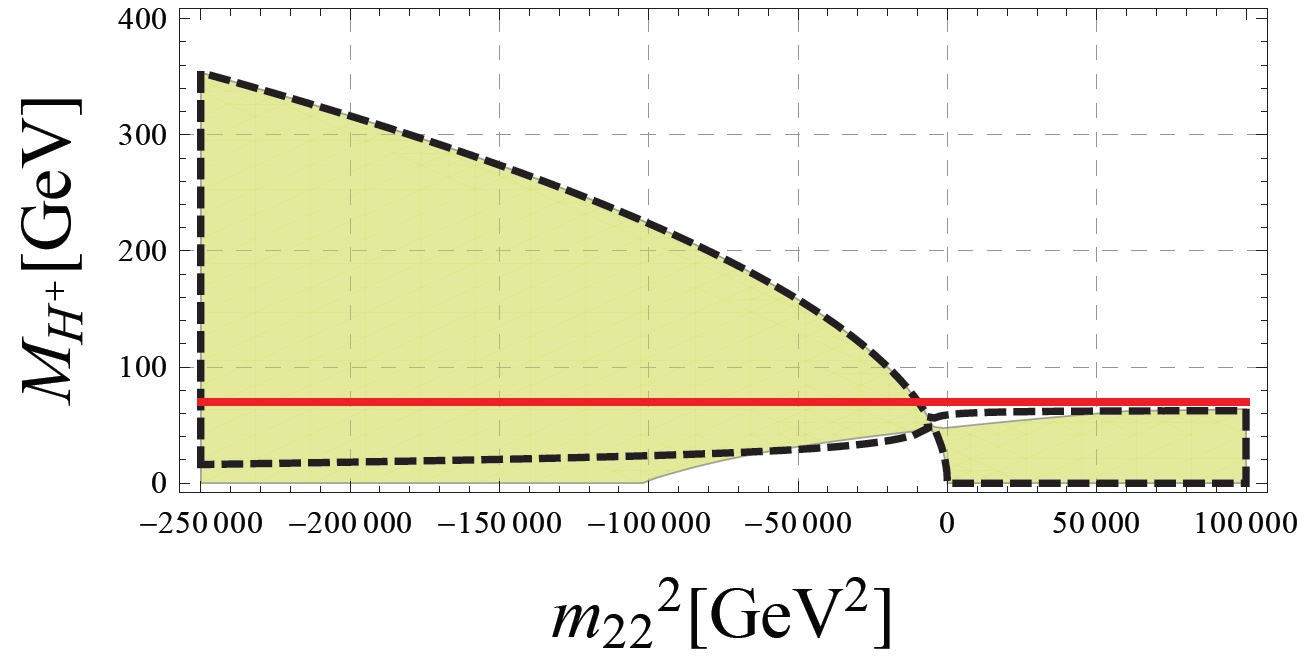}\\
\includegraphics[width=0.4\textwidth]{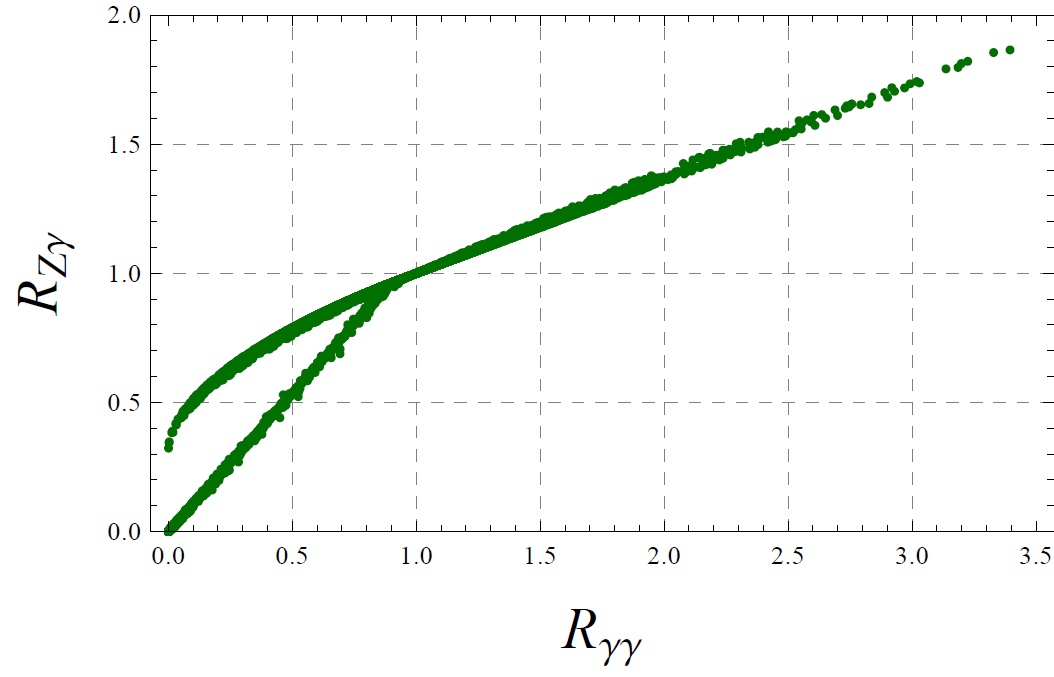}
\caption{Upper panel: Region in the $(m_{22}^2,\ \m)$ plane where $\rgt>1$ (shaded region) and the region where $\rzgt>1$ (inside the dashed line). Differences appear only below the $\m=70\g$ line (red solid line). Lower panel: Correlation between $\rg$ and $\rzg$.
\label{fig:rzg}}
\end{figure}

 Analogously to the $\rg$ case, $\rzg$ can be modified with respect to the SM case ($\rzg=1$) due to the charged scalar loop and the invisible decays. Just like in the $h\to \gamma\gamma$ case if the invisible decays of $h$ are open, $\rzg>1$ is impossible. The charged scalar loop is controlled by the $hH^+H^-$ coupling $\lc$ and the mass of the charged scalar (or equivalently $m_{22}^2$ and $\m$). Figure~\ref{fig:rzg} presents the region where $\rgt>1$ (shaded region) and the region where $\rzgt>1$ (inside the dashed line). It can be seen that the two regions overlap almost ideally, with differences only present for $\m<70\g$ (the red line corresponds to $\m=70\g$), i.e., in the region excluded by LEP. Thus, in the IDM if $\rg>1$ than also $\rzg>1$ and vice versa.

The same conclusion can be drawn from the plot in the lower panel of Fig.~\ref{fig:rzg}, where the correlation between $\rg$ and $\rzg$ is presented. The correlation is positive and the curve passes through the point $(1,1)$. The structure of a two-branch curve can be easily explained: the lower branch (straight line, for which $\rg\approx\rzg$) represents the case of open invisible channels, where both $\rg$ and $\rzg$ are damped by a  big common constant (invisible decays widths), which dominates over the charged scalar contributions, leading to $\rg\approx\rzg<1$. The other branch describes the correlation following from the fact that both $H^{\pm}$ loops in $h\to \gamma\gamma$ and $h\to Z \gamma$ are controlled by the same parameters.

We conclude that in the IDM the correlation between $\rg$ and $\rzg$ is positive and thus a measurement showing different result would exclude the IDM.
\section{Summary}\label{sum}

We  analyzed the diphoton decay rate of the Higgs boson in the IDM, for $M_h=125$ GeV, and presented a critical discussion of the results existing in the literature. The following conditions were taken into account: vacuum stability, existence of the Inert vacuum, perturbative unitarity, electroweak precision tests, and the LEP bounds. The importance of the condition determining the existence of the Inert vacuum should be emphasized as it significantly constrains the parameter space where $ \rg>1$ (i.e., excludes positive values of $m_{22}^2$). For the case of closed invisible decay channels of the Higgs boson we gave analytical solutions of the inequality $\rgt>1$ and for the general case we found the conditions using a random scan.

 We showed that the enhancement in the diphoton channel, with respect to the SM, is not possible if the invisible decay channels are open, confirming the results of Ref.~\cite{Arhrib:2012}. If the DM particle is heavier than $M_h/2$, then IDM can account for a Higgs boson with enhanced diphoton rate, while its remaining decay channels (apart from $h\to Z\gamma$), in particular the loop-induced decay $h\to gg$, stay SM-like.

The enhancement is only possible for $m_{22}^2<-9.8\cdot10^3\g^2$ ($\lc<0$) and the maximal $\rg$ value reaches 3.4. $\rg>1$ is possible for big values of $\m$, at the level of 1 TeV and higher, for large negative values of $m_{22}^2$.  However substantial enhancement can be realized only if the charged scalar is light.  $\rg>1.3$ implies  $\m\lesssim 135\g$,  and hence $62.4\g<M_H\lesssim135\g$, which would exclude the light and heavy DM scenarios. In this case stringent constraints on scalar couplings also arise: $-1.46<\lc,\lczp<-0.24$.

In the IDM the $Z\gamma$-decay rate is positively correlated with $\rg$, with a maximal value around 1.9. Regions where $\rg>1$ and where $\rzg>1$ overlap for $\m>70\g$, so $\rzg$ can be enhanced if and only if $\rg$ is enhanced.

\section*{Acknowledgments}
We would like to thank A.~Arhrib for discussion, materials, and pointing out an incorrect statement, and D.~Soko\l owska for valuable discussions.

\appendix
\section{Derivation of the solution of $\rg>1$ for $M_H>M_h/2$\label{wypr}}
We need to solve the following inequality for $M_h=125\g$
\be\label{nier}
|\mathcal{M}^{\textrm{SM}}+\delta\mathcal{M}^{\textrm{IDM}}|^2>|\mathcal{M}^{\textrm{SM}}|^2,
\ee
while $\mathcal{M}^{\textrm{SM}}$ is fixed. Let us use the following definitions: $a=\textrm{Re}\mathcal{M}^{\textrm{SM}}$, $b=\textrm{Im}\mathcal{M}^{\textrm{SM}}$ and $c=\delta\mathcal{M}^{\textrm{IDM}}=\frac{2\m^2+m_{22}^2}{2\m^2}A_0(\tau)$, where $\tau=\frac{4\m^2}{M_h^2}$, $\tau>1$. The parameter $c\in\mathbb{R}$, because $f\Big(\frac{4\m^2}{M_h^2}\Big)=\arcsin^2\big(\frac{M_h}{2M_H}\big)$ for $M_H>M_h/2$. Hence the inequality~(\ref{nier}) can be written as $|a+ib+c|^2>|a+ib|^2$ and is equivalent to $c(c+2a)>0$.

There are two possibilities:
$c>0$ and $c+2a>0$ or
$c<0$ and $c+2a<0$. One can compute that $a\approx - 6.53 <0$, so the two cases reduce to $c>-2a$ or $c<0$.

$c>-2a$ if and only if $\frac{2\m^2+m_{22}^2}{2\m^2}A_0(\tau)>-2a$. $A_0(\tau)=-\tau+\tau^2\arcsin^2(1/\tau)$, so $A_0(\tau)>0$ for $\tau>1$. 
Therefore we have
$$
m_{22}^2>\frac{a M_h^2}{1-\left(\frac{2 \m}{M_h}\right)^2\arcsin^2\left(\frac{M_h}{2 \m}\right)}-2\m^2.
$$

$c>0$ in two cases: either if  $2\m^2+m_{22}^2>0$ and $A_0(\tau)<0$, or if $2\m^2+m_{22}^2<0$ and $A_0(\tau)>0$.  As $A_0(\tau)>0$, the first option is excluded and the other reduces to $m_{22}^2<-2\m^2$.

Finally, there are two regions where enhancement in the $h\to\gamma\gamma$ channel is possible:
\begin{align}
&m_{22}^2<-2\m^2\quad
\textrm{or}\nonumber\\*[4pt]
&m_{22}^2>\frac{a M_h^2}{1-\left(\frac{2 \m}{M_h}\right)^2\arcsin^2\left(\frac{M_h}{2 \m}\right)}-2\m^2.\nonumber
\end{align}

\section{Decay widths of the Higgs boson\label{dec}}

Below we summarize the decay widths of the Higgs boson following~Refs.~\cite{Djouadi:2005sm, Djouadi:2005, Beringer:2012,Djouadi:1995, Chen:2013}.
\begin{enumerate}
\item $h\to q\bar{q}$
\begin{align}
\Gamma(h\to q\bar{q})=&\frac{3G_F}{4\sqrt{2}\pi}M_h \overline{m}_q^2(M_h)\Bigg\{1+5.67\frac{\overline{\alpha}_s(M_h)}{\pi}+\nonumber\\*
&+\bigg[37.51-1.36N_f-\frac{2}{3}\log\frac{M_h^2}{m_t^2}+\nonumber\\*
&+\bigg(\frac{1}{3}\log\frac{\overline{m}_q^2(M_h)}{M_h^2}\bigg)^2\bigg]\frac{\overline{\alpha}_s^2(M_h)}{\pi^2}\Bigg\}.\nonumber
\end{align}
$N_f=5$ is the number of active light-quark flavors. The running quark mass defined at the scale $M_h$ is~\cite{Djouadi:1995}
\begin{align}
\overline{m}_q(M_h)=&\overline{m}_q(m_q)\bigg(\frac{\overline{\alpha}_s(M_h)}{\overline{\alpha}_s(m_q)}\bigg)^{12/(33-2N_f)}\times\nonumber\\*
&\times\frac{1+c_{1q}\overline{\alpha}_s(M_h)/\pi+c_{2q}\overline{\alpha}_s^2(M_h)/\pi^2}{1+c_{1q}\overline{\alpha}_s(m_q)/\pi+c_{2q}\overline{\alpha}_s^2(m_q)/\pi^2},\nonumber
\end{align}
where for the bottom quark $c_{1b}=1.17$, $c_{2b}=1.50$ and for the charm quark $c_{1c}=1.01$, $c_{2c}=1.39$. The running strong coupling constant is approximated at the one-loop level (for energy scales around $M_h$, where the number of active light quarks can be taken  to be constant)~\cite{Beringer:2012}
$$
\overline{\alpha}_s(M_h)=\frac{\overline{\alpha}_s(M_Z)}{1+\frac{33-2N_f}{12\pi}\overline{\alpha}_s(M_Z)\log\frac{M_h^2}{M_Z^2}}.
$$
The values of quark masses and of the strong coupling  are taken from the Particle Data Group~\cite{Beringer:2012}: $\overline{m}_b(m_b)=4.18\g$, $\overline{m}_c(m_c)=1.273\g$, $\overline{\alpha}_s(M_Z)=0.118$, $\overline{\alpha}_s(m_b)=0.223$, and $\overline{\alpha}_s(m_c)=0.38$.

\item $h\to\tau^+\tau^-$ $$
\Gamma(h\to\tau^+\tau^-)=\frac{G_F N_c}{4\sqrt{2}\pi}M_h m_{\tau}^2 \bigg(1-\frac{4m_{\tau}^2}{M_h^2}\bigg)^{3/2}.
$$


\item $h\to VV^*$ $$
\Gamma(h\to VV^*)=\frac{3G_F^2}{16\pi^3}M_V^4M_h\delta_VR_T(x),
$$
where $\delta_W=1$, $\delta_Z=\frac{7}{12}-\frac{10}{9}\sin^2\theta_W+\frac{40}{9}\sin^4\theta_W$,
\begin{align}
R_T(x)=&\frac{3(1-8x+20x^2)}{\sqrt{4x-1}}\arccos\Big(\frac{3x-1}{2x^{3/2}}\Big)\nonumber\\*
&-\frac{1-x}{2x}\big(2-13x+47x^2\big)\nonumber\\*
&-\frac{3}{2}(1-6x+4x^2)\log x\nonumber
\end{align}
and $x=\frac{M_V^2}{M_h^2}$.


\item $h\to gg$ $$
\Gamma(h\to gg)=\frac{G_F\alpha_s^2M_h^3}{36\sqrt{2}\pi^3}\bigg|\frac{3}{4}A_{1/2}\bigg(\frac{4m_t^2}{M_h^2}\bigg)\bigg|^2.
$$

\item $h\to \varphi\varphi$ ($\varphi=H,A$)
$$
\Gamma(h\to \varphi\varphi)=\frac{\lambda_{h\varphi\varphi}^2v^2}{32\pi M_h}\sqrt{1-\frac{4M_{\varphi}^2}{M_h^2}},
$$
where $\lambda_{hHH}=\lambda_{345}$ and $\lambda_{hAA}=\lambda_{345}^-$.

\begin{widetext}
\item $h\to Z\gamma$
\begin{align}
\Gamma(h\to Z\gamma)= &\frac{G_F^2\alpha}{64\pi^4} M_W^2 M_h^3 \bigg(1-\frac{M_Z^2}{M_h^2}\bigg)^3
\bigg| 2\frac{1-\frac{8}{3}\sin^2\theta_W}{\cos\theta_W}
A_{1/2}^h\bigg(\frac{4m_t^2}{M_h^2},\frac{4m_t^2}{M_Z^2}\bigg)
+A_1^h\bigg(\frac{4M_W^2}{M_h^2},\frac{4M_W^2}{M_Z^2}\bigg)\nonumber\\*
&-\frac{2\m^2+m_{22}^2}{2\m^2}\frac{(1-2\sin^2\theta_W)}{\cos\theta_W}I_1\left(\frac{4\m^2}{M_h^2},\frac{4\m^2}{M_Z^2}\right)\bigg|^2,\label{rzg-wzor}
\end{align} where
\begin{eqnarray}
A_{1/2}^h(\tau,\lambda)&=&I_1(\tau,\lambda)-I_2(\tau,\lambda),\nonumber\\*
A_1^h(\tau,\lambda)&=&\cos\theta_W\Bigg\{4\bigg(3-\frac{\sin^2\theta_W}{\cos^2\theta_W}\bigg)I_2(\tau,\lambda)+\bigg[\bigg(1+\frac{2}{\tau}\bigg)\frac{\sin^2\theta_W}{\cos^2\theta_W}-\bigg(5+\frac{2}{\tau}\bigg)\bigg]I_1(\tau,\lambda)\Bigg\},\nonumber\\*
I_1(\tau,\lambda)&=&\frac{\tau\lambda}{2(\tau-\lambda)}+\frac{\tau^2\lambda^2}{2(\tau-\lambda)^2}\big[f(\tau)-f(\lambda)\big]+\frac{\tau^2\lambda}{(\tau-\lambda)^2}\big[g(\tau^{-1})-g(\lambda^{-1})\big],\nonumber\\*
I_2(\tau,\lambda)&=&-\frac{\tau\lambda}{2(\tau-\lambda)}\big[f(\tau)-f(\lambda)\big],\nonumber
\end{eqnarray}
$$
g(\tau)=\begin{cases}
\sqrt{\frac{1}{\tau}-1}\arcsin\sqrt{\tau}& \textrm{for }\tau\leqslant 1,\\[4pt]
\frac{\sqrt{1-\frac{1}{\tau}}}{2}\left(\log\frac{1+\sqrt{1-\frac{1}{\tau}}}{1-\sqrt{1-\frac{1}{\tau}}}-i\pi\right)&\textrm{for } \tau>1.
\end{cases}
$$
Note the minus sign of the charged scalar contribution~\cite{Chen:2013}, which is different than the result in Ref.~\cite{Djouadi:2005}.
\end{widetext}
\end{enumerate}


\vspace{\stretch{1}}

\bibliography{biblio}

\begin{thebibliography}{31}%
\makeatletter
\providecommand \@ifxundefined [1]{%
 \@ifx{#1\undefined}
}%
\providecommand \@ifnum [1]{%
 \ifnum #1\expandafter \@firstoftwo
 \else \expandafter \@secondoftwo
 \fi
}%
\providecommand \@ifx [1]{%
 \ifx #1\expandafter \@firstoftwo
 \else \expandafter \@secondoftwo
 \fi
}%
\providecommand \natexlab [1]{#1}%
\providecommand \enquote  [1]{``#1''}%
\providecommand \bibnamefont  [1]{#1}%
\providecommand \bibfnamefont [1]{#1}%
\providecommand \citenamefont [1]{#1}%
\providecommand \href@noop [0]{\@secondoftwo}%
\providecommand \href [0]{\begingroup \@sanitize@url \@href}%
\providecommand \@href[1]{\@@startlink{#1}\@@href}%
\providecommand \@@href[1]{\endgroup#1\@@endlink}%
\providecommand \@sanitize@url [0]{\catcode `\\12\catcode `\$12\catcode
  `\&12\catcode `\#12\catcode `\^12\catcode `\_12\catcode `\%12\relax}%
\providecommand \@@startlink[1]{}%
\providecommand \@@endlink[0]{}%
\providecommand \url  [0]{\begingroup\@sanitize@url \@url }%
\providecommand \@url [1]{\endgroup\@href {#1}{\urlprefix }}%
\providecommand \urlprefix  [0]{URL }%
\providecommand \Eprint [0]{\href }%
\providecommand \doibase [0]{http://dx.doi.org/}%
\providecommand \selectlanguage [0]{\@gobble}%
\providecommand \bibinfo  [0]{\@secondoftwo}%
\providecommand \bibfield  [0]{\@secondoftwo}%
\providecommand \translation [1]{[#1]}%
\providecommand \BibitemOpen [0]{}%
\providecommand \bibitemStop [0]{}%
\providecommand \bibitemNoStop [0]{.\EOS\space}%
\providecommand \EOS [0]{\spacefactor3000\relax}%
\providecommand \BibitemShut  [1]{\csname bibitem#1\endcsname}%
\let\auto@bib@innerbib\@empty
\bibitem [{\citenamefont {Aad}\ \emph {et~al.}(2012)\citenamefont {Aad} \emph
  {et~al.}}]{atlas:2012}%
  \BibitemOpen
  \bibfield  {author} {\bibinfo {author} {\bibfnamefont {G.}~\bibnamefont
  {Aad}} \emph {et~al.} (\bibinfo {collaboration} {ATLAS Collaboration}),\
  }\href {\doibase 10.1016/j.physletb.2012.08.020} {\bibfield  {journal}
  {\bibinfo  {journal} {Phys.Lett.}\ }\textbf {\bibinfo {volume} {B716}},\
  \bibinfo {pages} {1} (\bibinfo {year} {2012})},\ \Eprint
  {http://arxiv.org/abs/1207.7214} {arXiv:1207.7214 [hep-ex]} \BibitemShut
  {NoStop}%
\bibitem [{\citenamefont {Chatrchyan}\ \emph {et~al.}(2012)\citenamefont
  {Chatrchyan} \emph {et~al.}}]{cms:2012}%
  \BibitemOpen
  \bibfield  {author} {\bibinfo {author} {\bibfnamefont {S.}~\bibnamefont
  {Chatrchyan}} \emph {et~al.} (\bibinfo {collaboration} {CMS Collaboration}),\
  }\href {\doibase 10.1016/j.physletb.2012.08.021} {\bibfield  {journal}
  {\bibinfo  {journal} {Phys.Lett.}\ }\textbf {\bibinfo {volume} {B716}},\
  \bibinfo {pages} {30} (\bibinfo {year} {2012})},\ \Eprint
  {http://arxiv.org/abs/1207.7235} {arXiv:1207.7235 [hep-ex]} \BibitemShut
  {NoStop}%
\bibitem [{\citenamefont {Aad}\ \emph {et~al.}()\citenamefont {Aad} \emph
  {et~al.}}]{Atlas:12-2012}%
  \BibitemOpen
  \bibfield  {author} {\bibinfo {author} {\bibfnamefont {G.}~\bibnamefont
  {Aad}} \emph {et~al.} (\bibinfo {collaboration} {ATLAS Collaboration}),\
  }\href@noop {} {\bibfield  {journal} {\bibinfo  {journal} {ATLAS NOTE}\
  }\textbf {\bibinfo {volume} {ATLAS-CONF-2012-168}}}\BibitemShut {NoStop}%
\bibitem [{\citenamefont {Cao}\ \emph {et~al.}(2007)\citenamefont {Cao},
  \citenamefont {Ma},\ and\ \citenamefont {Rajasekaran}}]{Ma:2007}%
  \BibitemOpen
  \bibfield  {author} {\bibinfo {author} {\bibfnamefont {Q.-H.}\ \bibnamefont
  {Cao}}, \bibinfo {author} {\bibfnamefont {E.}~\bibnamefont {Ma}}, \ and\
  \bibinfo {author} {\bibfnamefont {G.}~\bibnamefont {Rajasekaran}},\ }\href
  {\doibase 10.1103/PhysRevD.76.095011} {\bibfield  {journal} {\bibinfo
  {journal} {Phys.Rev.}\ }\textbf {\bibinfo {volume} {D76}},\ \bibinfo {pages}
  {095011} (\bibinfo {year} {2007})},\ \Eprint {http://arxiv.org/abs/0708.2939}
  {arXiv:0708.2939 [hep-ph]} \BibitemShut {NoStop}%
\bibitem [{\citenamefont {Posch}(2011)}]{Posch:2010}%
  \BibitemOpen
  \bibfield  {author} {\bibinfo {author} {\bibfnamefont {P.}~\bibnamefont
  {Posch}},\ }\href {\doibase 10.1016/j.physletb.2011.01.003} {\bibfield
  {journal} {\bibinfo  {journal} {Phys.Lett.}\ }\textbf {\bibinfo {volume}
  {B696}},\ \bibinfo {pages} {447} (\bibinfo {year} {2011})},\ \Eprint
  {http://arxiv.org/abs/1001.1759} {arXiv:1001.1759 [hep-ph]} \BibitemShut
  {NoStop}%
\bibitem [{\citenamefont {Arhrib}\ \emph {et~al.}(2012)\citenamefont {Arhrib},
  \citenamefont {Benbrik},\ and\ \citenamefont {Gaur}}]{Arhrib:2012}%
  \BibitemOpen
  \bibfield  {author} {\bibinfo {author} {\bibfnamefont {A.}~\bibnamefont
  {Arhrib}}, \bibinfo {author} {\bibfnamefont {R.}~\bibnamefont {Benbrik}}, \
  and\ \bibinfo {author} {\bibfnamefont {N.}~\bibnamefont {Gaur}},\ }\href
  {\doibase 10.1103/PhysRevD.85.095021} {\bibfield  {journal} {\bibinfo
  {journal} {Phys.Rev.}\ }\textbf {\bibinfo {volume} {D85}},\ \bibinfo {pages}
  {095021} (\bibinfo {year} {2012})},\ \Eprint {http://arxiv.org/abs/1201.2644}
  {arXiv:1201.2644 [hep-ph]} \BibitemShut {NoStop}%
\bibitem [{\citenamefont {Chang}\ \emph {et~al.}(2012)\citenamefont {Chang},
  \citenamefont {Cheung}, \citenamefont {Tseng},\ and\ \citenamefont
  {Yuan}}]{Chang:2012}%
  \BibitemOpen
  \bibfield  {author} {\bibinfo {author} {\bibfnamefont {J.}~\bibnamefont
  {Chang}}, \bibinfo {author} {\bibfnamefont {K.}~\bibnamefont {Cheung}},
  \bibinfo {author} {\bibfnamefont {P.-Y.}\ \bibnamefont {Tseng}}, \ and\
  \bibinfo {author} {\bibfnamefont {T.-C.}\ \bibnamefont {Yuan}},\ }\href
  {\doibase 10.1142/S0217751X1230030X} {\bibfield  {journal} {\bibinfo
  {journal} {Int.J.Mod.Phys.}\ }\textbf {\bibinfo {volume} {A27}},\ \bibinfo
  {pages} {1230030} (\bibinfo {year} {2012})},\ \Eprint
  {http://arxiv.org/abs/1211.6823} {arXiv:1211.6823 [hep-ph]} \BibitemShut
  {NoStop}%
\bibitem [{\citenamefont {Borah}\ and\ \citenamefont
  {Cline}(2012)}]{Cline:2012}%
  \BibitemOpen
  \bibfield  {author} {\bibinfo {author} {\bibfnamefont {D.}~\bibnamefont
  {Borah}}\ and\ \bibinfo {author} {\bibfnamefont {J.~M.}\ \bibnamefont
  {Cline}},\ }\href {\doibase 10.1103/PhysRevD.86.055001} {\bibfield  {journal}
  {\bibinfo  {journal} {Phys.Rev.}\ }\textbf {\bibinfo {volume} {D86}},\
  \bibinfo {pages} {055001} (\bibinfo {year} {2012})},\ \Eprint
  {http://arxiv.org/abs/1204.4722} {arXiv:1204.4722 [hep-ph]} \BibitemShut
  {NoStop}%
\bibitem [{\citenamefont {Deshpande}\ and\ \citenamefont {Ma}(1978)}]{Ma:1978}%
  \BibitemOpen
  \bibfield  {author} {\bibinfo {author} {\bibfnamefont {N.~G.}\ \bibnamefont
  {Deshpande}}\ and\ \bibinfo {author} {\bibfnamefont {E.}~\bibnamefont {Ma}},\
  }\href {\doibase 10.1103/PhysRevD.18.2574} {\bibfield  {journal} {\bibinfo
  {journal} {Phys.Rev.}\ }\textbf {\bibinfo {volume} {D18}},\ \bibinfo {pages}
  {2574} (\bibinfo {year} {1978})}\BibitemShut {NoStop}%
\bibitem [{\citenamefont {Barbieri}\ \emph {et~al.}(2006)\citenamefont
  {Barbieri}, \citenamefont {Hall},\ and\ \citenamefont
  {Rychkov}}]{Barbieri:2006}%
  \BibitemOpen
  \bibfield  {author} {\bibinfo {author} {\bibfnamefont {R.}~\bibnamefont
  {Barbieri}}, \bibinfo {author} {\bibfnamefont {L.~J.}\ \bibnamefont {Hall}},
  \ and\ \bibinfo {author} {\bibfnamefont {V.~S.}\ \bibnamefont {Rychkov}},\
  }\href {\doibase 10.1103/PhysRevD.74.015007} {\bibfield  {journal} {\bibinfo
  {journal} {Phys.Rev.}\ }\textbf {\bibinfo {volume} {D74}},\ \bibinfo {pages}
  {015007} (\bibinfo {year} {2006})},\ \Eprint
  {http://arxiv.org/abs/hep-ph/0603188} {arXiv:hep-ph/0603188 [hep-ph]}
  \BibitemShut {NoStop}%
\bibitem [{\citenamefont {Ginzburg}\ \emph {et~al.}(2010)\citenamefont
  {Ginzburg}, \citenamefont {Kanishev}, \citenamefont {Krawczyk},\ and\
  \citenamefont {Sokołowska}}]{Krawczyk:2010}%
  \BibitemOpen
  \bibfield  {author} {\bibinfo {author} {\bibfnamefont {I.}~\bibnamefont
  {Ginzburg}}, \bibinfo {author} {\bibfnamefont {K.}~\bibnamefont {Kanishev}},
  \bibinfo {author} {\bibfnamefont {M.}~\bibnamefont {Krawczyk}}, \ and\
  \bibinfo {author} {\bibfnamefont {D.}~\bibnamefont {Sokołowska}},\ }\href
  {\doibase 10.1103/PhysRevD.82.123533} {\bibfield  {journal} {\bibinfo
  {journal} {Phys.Rev.}\ }\textbf {\bibinfo {volume} {D82}},\ \bibinfo {pages}
  {123533} (\bibinfo {year} {2010})},\ \Eprint {http://arxiv.org/abs/1009.4593}
  {arXiv:1009.4593 [hep-ph]} \BibitemShut {NoStop}%
\bibitem [{\citenamefont {Ginzburg}\ and\ \citenamefont
  {Krawczyk}(2005)}]{Krawczyk:2004sym}%
  \BibitemOpen
  \bibfield  {author} {\bibinfo {author} {\bibfnamefont {I.~F.}\ \bibnamefont
  {Ginzburg}}\ and\ \bibinfo {author} {\bibfnamefont {M.}~\bibnamefont
  {Krawczyk}},\ }\href {\doibase 10.1103/PhysRevD.72.115013} {\bibfield
  {journal} {\bibinfo  {journal} {Phys.Rev.}\ }\textbf {\bibinfo {volume}
  {D72}},\ \bibinfo {pages} {115013} (\bibinfo {year} {2005})},\ \Eprint
  {http://arxiv.org/abs/hep-ph/0408011} {arXiv:hep-ph/0408011 [hep-ph]}
  \BibitemShut {NoStop}%
\bibitem [{\citenamefont {Branco}\ \emph {et~al.}(1999)\citenamefont {Branco},
  \citenamefont {Lavoura},\ and\ \citenamefont {Silva}}]{Branco:1999}%
  \BibitemOpen
  \bibfield  {author} {\bibinfo {author} {\bibfnamefont {G.~C.}\ \bibnamefont
  {Branco}}, \bibinfo {author} {\bibfnamefont {L.}~\bibnamefont {Lavoura}}, \
  and\ \bibinfo {author} {\bibfnamefont {J.~P.}\ \bibnamefont {Silva}},\
  }\href@noop {} {\emph {\bibinfo {title} {{CP Violation}}}}\ (\bibinfo
  {publisher} {Oxford University Press},\ \bibinfo {year} {1999})\BibitemShut
  {NoStop}%
\bibitem [{\citenamefont {Dolle}\ and\ \citenamefont {Su}(2009)}]{Dolle:2009}%
  \BibitemOpen
  \bibfield  {author} {\bibinfo {author} {\bibfnamefont {E.~M.}\ \bibnamefont
  {Dolle}}\ and\ \bibinfo {author} {\bibfnamefont {S.}~\bibnamefont {Su}},\
  }\href {\doibase 10.1103/PhysRevD.80.055012} {\bibfield  {journal} {\bibinfo
  {journal} {Phys.Rev.}\ }\textbf {\bibinfo {volume} {D80}},\ \bibinfo {pages}
  {055012} (\bibinfo {year} {2009})},\ \Eprint {http://arxiv.org/abs/0906.1609}
  {arXiv:0906.1609 [hep-ph]} \BibitemShut {NoStop}%
\bibitem [{\citenamefont {Lopez~Honorez}\ \emph {et~al.}(2007)\citenamefont
  {Lopez~Honorez}, \citenamefont {Nezri}, \citenamefont {Oliver},\ and\
  \citenamefont {Tytgat}}]{LopezHonorez:2006}%
  \BibitemOpen
  \bibfield  {author} {\bibinfo {author} {\bibfnamefont {L.}~\bibnamefont
  {Lopez~Honorez}}, \bibinfo {author} {\bibfnamefont {E.}~\bibnamefont
  {Nezri}}, \bibinfo {author} {\bibfnamefont {J.~F.}\ \bibnamefont {Oliver}}, \
  and\ \bibinfo {author} {\bibfnamefont {M.~H.}\ \bibnamefont {Tytgat}},\
  }\href {\doibase 10.1088/1475-7516/2007/02/028} {\bibfield  {journal}
  {\bibinfo  {journal} {JCAP}\ }\textbf {\bibinfo {volume} {0702}},\ \bibinfo
  {pages} {028} (\bibinfo {year} {2007})},\ \Eprint
  {http://arxiv.org/abs/hep-ph/0612275} {arXiv:hep-ph/0612275 [hep-ph]}
  \BibitemShut {NoStop}%
\bibitem [{\citenamefont {Lopez~Honorez}(2007)}]{LopezHonorez:2007}%
  \BibitemOpen
  \bibfield  {author} {\bibinfo {author} {\bibfnamefont {L.}~\bibnamefont
  {Lopez~Honorez}},\ }\href@noop {} {\  (\bibinfo {year} {2007})},\ \Eprint
  {http://arxiv.org/abs/0706.0186} {arXiv:0706.0186 [hep-ph]} \BibitemShut
  {NoStop}%
\bibitem [{\citenamefont {Sokołowska}(2011)}]{Sokolowska:2011}%
  \BibitemOpen
  \bibfield  {author} {\bibinfo {author} {\bibfnamefont {D.}~\bibnamefont
  {Sokołowska}},\ }\href@noop {} {\  (\bibinfo {year} {2011})},\ \Eprint
  {http://arxiv.org/abs/1107.1991} {arXiv:1107.1991 [hep-ph]} \BibitemShut
  {NoStop}%
\bibitem [{\citenamefont {Swiezewska}(2012)}]{Swiezewska:2012}%
  \BibitemOpen
  \bibfield  {author} {\bibinfo {author} {\bibfnamefont {B.}~\bibnamefont
  {Swiezewska}},\ }\href@noop {} {\  (\bibinfo {year} {2012})},\ \Eprint
  {http://arxiv.org/abs/1209.5725} {arXiv:1209.5725 [hep-ph]} \BibitemShut
  {NoStop}%
\bibitem [{\citenamefont {Kanemura}\ \emph {et~al.}(1993)\citenamefont
  {Kanemura}, \citenamefont {Kubota},\ and\ \citenamefont
  {Takasugi}}]{Kanemura:1993}%
  \BibitemOpen
  \bibfield  {author} {\bibinfo {author} {\bibfnamefont {S.}~\bibnamefont
  {Kanemura}}, \bibinfo {author} {\bibfnamefont {T.}~\bibnamefont {Kubota}}, \
  and\ \bibinfo {author} {\bibfnamefont {E.}~\bibnamefont {Takasugi}},\ }\href
  {\doibase 10.1016/0370-2693(93)91205-2} {\bibfield  {journal} {\bibinfo
  {journal} {Phys.Lett.}\ }\textbf {\bibinfo {volume} {B313}},\ \bibinfo
  {pages} {155} (\bibinfo {year} {1993})},\ \Eprint
  {http://arxiv.org/abs/hep-ph/9303263} {arXiv:hep-ph/9303263 [hep-ph]}
  \BibitemShut {NoStop}%
\bibitem [{\citenamefont {Akeroyd}\ \emph {et~al.}(2000)\citenamefont
  {Akeroyd}, \citenamefont {Arhrib},\ and\ \citenamefont
  {Naimi}}]{Akeroyd:2000}%
  \BibitemOpen
  \bibfield  {author} {\bibinfo {author} {\bibfnamefont {A.~G.}\ \bibnamefont
  {Akeroyd}}, \bibinfo {author} {\bibfnamefont {A.}~\bibnamefont {Arhrib}}, \
  and\ \bibinfo {author} {\bibfnamefont {E.-M.}\ \bibnamefont {Naimi}},\ }\href
  {\doibase 10.1016/S0370-2693(00)00962-X} {\bibfield  {journal} {\bibinfo
  {journal} {Phys.Lett.}\ }\textbf {\bibinfo {volume} {B490}},\ \bibinfo
  {pages} {119} (\bibinfo {year} {2000})},\ \Eprint
  {http://arxiv.org/abs/hep-ph/0006035} {arXiv:hep-ph/0006035 [hep-ph]}
  \BibitemShut {NoStop}%
\bibitem [{\citenamefont {Gorczyca}(2011)}]{praca-mag}%
  \BibitemOpen
  \bibfield  {author} {\bibinfo {author} {\bibfnamefont {B.}~\bibnamefont
  {Gorczyca}},\ }\href@noop {} {\enquote {\bibinfo {title} {{Unitarity
  constraints for the Inert Doublet Model (in Polish)}},}\ }\bibinfo
  {howpublished} {{Master Thesis at the University of Warsaw}} (\bibinfo {year}
  {2011})\BibitemShut {NoStop}%
\bibitem [{\citenamefont {Nakamura}\ \emph {et~al.}(2010)\citenamefont
  {Nakamura} \emph {et~al.}}]{Nakamura:2010}%
  \BibitemOpen
  \bibfield  {author} {\bibinfo {author} {\bibfnamefont {K.}~\bibnamefont
  {Nakamura}} \emph {et~al.} (\bibinfo {collaboration} {Particle Data Group}),\
  }\href {\doibase 10.1088/0954-3899/37/7A/075021} {\bibfield  {journal}
  {\bibinfo  {journal} {J.Phys.G}\ }\textbf {\bibinfo {volume} {G37}},\
  \bibinfo {pages} {075021} (\bibinfo {year} {2010})}\BibitemShut {NoStop}%
\bibitem [{\citenamefont {Lundstrom}\ \emph {et~al.}(2009)\citenamefont
  {Lundstrom}, \citenamefont {Gustafsson},\ and\ \citenamefont
  {Edsjo}}]{Gustafsson:2009}%
  \BibitemOpen
  \bibfield  {author} {\bibinfo {author} {\bibfnamefont {E.}~\bibnamefont
  {Lundstrom}}, \bibinfo {author} {\bibfnamefont {M.}~\bibnamefont
  {Gustafsson}}, \ and\ \bibinfo {author} {\bibfnamefont {J.}~\bibnamefont
  {Edsjo}},\ }\href {\doibase 10.1103/PhysRevD.79.035013} {\bibfield  {journal}
  {\bibinfo  {journal} {Phys.Rev.}\ }\textbf {\bibinfo {volume} {D79}},\
  \bibinfo {pages} {035013} (\bibinfo {year} {2009})},\ \Eprint
  {http://arxiv.org/abs/0810.3924} {arXiv:0810.3924 [hep-ph]} \BibitemShut
  {NoStop}%
\bibitem [{\citenamefont {Gustafsson}(2010)}]{Gustafsson:2010}%
  \BibitemOpen
  \bibfield  {author} {\bibinfo {author} {\bibfnamefont {M.}~\bibnamefont
  {Gustafsson}},\ }\href@noop {} {\bibfield  {journal} {\bibinfo  {journal}
  {PoS}\ }\textbf {\bibinfo {volume} {CHARGED2010}},\ \bibinfo {pages} {030}
  (\bibinfo {year} {2010})},\ \Eprint {http://arxiv.org/abs/1106.1719}
  {arXiv:1106.1719 [hep-ph]} \BibitemShut {NoStop}%
\bibitem [{\citenamefont {Djouadi}(2008{\natexlab{a}})}]{Djouadi:2005}%
  \BibitemOpen
  \bibfield  {author} {\bibinfo {author} {\bibfnamefont {A.}~\bibnamefont
  {Djouadi}},\ }\href {\doibase 10.1016/j.physrep.2007.10.005} {\bibfield
  {journal} {\bibinfo  {journal} {Phys.Rept.}\ }\textbf {\bibinfo {volume}
  {459}},\ \bibinfo {pages} {1} (\bibinfo {year} {2008}{\natexlab{a}})},\
  \Eprint {http://arxiv.org/abs/hep-ph/0503173} {arXiv:hep-ph/0503173 [hep-ph]}
  \BibitemShut {NoStop}%
\bibitem [{\citenamefont {Djouadi}(2008{\natexlab{b}})}]{Djouadi:2005sm}%
  \BibitemOpen
  \bibfield  {author} {\bibinfo {author} {\bibfnamefont {A.}~\bibnamefont
  {Djouadi}},\ }\href {\doibase 10.1016/j.physrep.2007.10.004} {\bibfield
  {journal} {\bibinfo  {journal} {Phys.Rept.}\ }\textbf {\bibinfo {volume}
  {457}},\ \bibinfo {pages} {1} (\bibinfo {year} {2008}{\natexlab{b}})},\
  \Eprint {http://arxiv.org/abs/hep-ph/0503172} {arXiv:hep-ph/0503172 [hep-ph]}
  \BibitemShut {NoStop}%
\bibitem [{\citenamefont {Djouadi}\ \emph {et~al.}(1996)\citenamefont
  {Djouadi}, \citenamefont {Kalinowski},\ and\ \citenamefont
  {Zerwas}}]{Djouadi:1995}%
  \BibitemOpen
  \bibfield  {author} {\bibinfo {author} {\bibfnamefont {A.}~\bibnamefont
  {Djouadi}}, \bibinfo {author} {\bibfnamefont {J.}~\bibnamefont {Kalinowski}},
  \ and\ \bibinfo {author} {\bibfnamefont {P.}~\bibnamefont {Zerwas}},\ }\href
  {\doibase 10.1007/s002880050121} {\bibfield  {journal} {\bibinfo  {journal}
  {Z.Phys.}\ }\textbf {\bibinfo {volume} {C70}},\ \bibinfo {pages} {435}
  (\bibinfo {year} {1996})},\ \Eprint {http://arxiv.org/abs/hep-ph/9511342}
  {arXiv:hep-ph/9511342 [hep-ph]} \BibitemShut {NoStop}%
\bibitem [{\citenamefont {Beringer}\ \emph {et~al.}(2012)\citenamefont
  {Beringer} \emph {et~al.}}]{Beringer:2012}%
  \BibitemOpen
  \bibfield  {author} {\bibinfo {author} {\bibfnamefont {J.}~\bibnamefont
  {Beringer}} \emph {et~al.} (\bibinfo {collaboration} {Particle Data Group}),\
  }\href {\doibase 10.1103/PhysRevD.86.010001} {\bibfield  {journal} {\bibinfo
  {journal} {Phys. Rev. D}\ }\textbf {\bibinfo {volume} {86}},\ \bibinfo
  {pages} {010001} (\bibinfo {year} {2012})}\BibitemShut {NoStop}%
\bibitem [{\citenamefont {Chen}\ \emph {et~al.}(2013)\citenamefont {Chen},
  \citenamefont {Geng}, \citenamefont {Huang},\ and\ \citenamefont
  {Tsai}}]{Chen:2013}%
  \BibitemOpen
  \bibfield  {author} {\bibinfo {author} {\bibfnamefont {C.-S.}\ \bibnamefont
  {Chen}}, \bibinfo {author} {\bibfnamefont {C.-Q.}\ \bibnamefont {Geng}},
  \bibinfo {author} {\bibfnamefont {D.}~\bibnamefont {Huang}}, \ and\ \bibinfo
  {author} {\bibfnamefont {L.-H.}\ \bibnamefont {Tsai}},\ }\href {\doibase
  10.1103/PhysRevD.87.075019} {\bibfield  {journal} {\bibinfo  {journal} {Phys.
  Rev. D87,}\ }\textbf {\bibinfo {volume} {075019}} (\bibinfo {year} {2013}),\
  10.1103/PhysRevD.87.075019},\ \Eprint {http://arxiv.org/abs/1301.4694}
  {arXiv:1301.4694 [hep-ph]} \BibitemShut {NoStop}%
\bibitem [{\citenamefont {Spira}(1998)}]{Spira:1997}%
  \BibitemOpen
  \bibfield  {author} {\bibinfo {author} {\bibfnamefont {M.}~\bibnamefont
  {Spira}},\ }\href@noop {} {\bibfield  {journal} {\bibinfo  {journal}
  {Fortsch.Phys.}\ }\textbf {\bibinfo {volume} {46}},\ \bibinfo {pages} {203}
  (\bibinfo {year} {1998})},\ \Eprint {http://arxiv.org/abs/hep-ph/9705337}
  {arXiv:hep-ph/9705337 [hep-ph]} \BibitemShut {NoStop}%
\bibitem [{\citenamefont {Krawczyk}\ \emph {et~al.}(2013)\citenamefont
  {Krawczyk}, \citenamefont {Sokolowska}, \citenamefont {Swaczyna},\ and\
  \citenamefont {Swiezewska}}]{Krawczyk:2013}%
  \BibitemOpen
  \bibfield  {author} {\bibinfo {author} {\bibfnamefont {M.}~\bibnamefont
  {Krawczyk}}, \bibinfo {author} {\bibfnamefont {D.}~\bibnamefont
  {Sokolowska}}, \bibinfo {author} {\bibfnamefont {P.}~\bibnamefont
  {Swaczyna}}, \ and\ \bibinfo {author} {\bibfnamefont {B.}~\bibnamefont
  {Swiezewska}},\ }\href@noop {} {\  (\bibinfo {year} {2013})},\ \Eprint
  {http://arxiv.org/abs/1305.6266} {arXiv:1305.6266 [hep-ph]} \BibitemShut
  {NoStop}%
\end{thebibliography}%

\end{document}